\documentclass[11pt]{article}

\makeatletter
\@addtoreset{equation}{section}
\makeatother

\newlength{\dinwidth}                       
\newlength{\dinmargin}                      
\def\Journal#1#2#3#4{{#1} {\bf #2}, #3 (#4)}

\def\NCA{\em Nuovo Cimento}

\def\NPB{{\em Nucl. Phys.} B}
\def\PLB{{\em Phys. Lett.}  B}
\def\PRL{\em Phys. Rev. Lett.}
\def\PR{\em Phys. Rev.}
\def\PRD{{\em Phys. Rev.} D}

\def\be{\begin{equation}}
\def\ee{\end{equation}}
\def\bea{\begin{eqnarray}}
\def\eea{\end{eqnarray}}

\begin{document}

\thispagestyle{empty}
\def\thefootnote{\fnsymbol{footnote}}
\setcounter{footnote}{1}

\hfill \begin{minipage}[t]{4cm}BI-TP 96/40\\
August 1996  
\end{minipage}

\vspace*{2cm}

\begin{center}
{\Large\bf ANOMALOUS ELECTROMAGNETIC PROCESSES\\
        AT HIGH TEMPERATURES \footnote{
       Invited talk at the Workshop on Quantum Chromodynamics:
       Confinement, Collisions, and Chaos,
       3 - 8 June  1996, The American University of Paris, Paris}
}

\end{center}

\vspace*{0.3cm}
\begin{center}
{\bf R.~Baier}$^1$, {\bf M.~Dirks}$^1$  {\bf and O.~Kober}$^1$
\\[.3cm]
$^1${\it Fakult\"at f\"ur Physik, Universit\"at Bielefeld, 
D-33501 Bielefeld, Germany}\\

\end{center}

\vspace*{3cm}
\section*{Abstract}
Chiral theories of constituent quarks interacting
with bosons and photons at high temperatures are studied.
In the expected chirally symmetric phase 
 effective electromagnetic anomalous couplings for e.g.
$\pi \sigma \to \gamma \gamma, ~ 
\gamma \to \pi \pi \pi \sigma, $  etc.,
are derived by applying functional methods.

\vfill

\def\thefootnote{\arabic{footnote}}
\setcounter{footnote}{0}
\clearpage

\section{Introduction}

In this contribution we discuss our approach~\cite{baier}
to confirm the interesting conjecture by Pisarski~\cite{pisa1,pisa2},
 who pointed out that although
the chiral (abelian as well as nonabelian) anomaly~\cite{ra,rb}
in terms of fundamental fields is temperature independent~\cite{rc},
the manifestation of the anomaly, however, in terms of effective fields 
changes with temperature.
When considering e.g. $\pi^0 \to 2 \gamma$ the observation~\cite{pisa1,pisa2}
is: "In a hot, chirally symmmetric phase, $\pi^0$ doesn't go into $2 \gamma$,
but $\pi^0 \sigma$ does"!

In the following we restrict our discussion to these electromagnetic
couplings, which  have first been found~\cite{pisa1,pisa2} in the framework
 of the constituent quark model~\cite{donoh}
by calculating the contribution from the Feynman one-loop
triangle (box) diagrams  at nonzero,
high temperature $T$.

Our approach also starts from the linear sigma model with
constituent quarks interacting with bosons and photons~\cite{donoh}.
In order to include the effect of the axial anomaly on the 
bosons we transform  the basis of right- and left-handed
quarks following Manohar et al.~\cite{manm}, but 
we take into account 
a chirally symmetric phase~\cite{rj} at high $T$.
 At $T = 0$ and in
the spontaneously broken phase this procedure gives
 the anomalous couplings by 
 applying functional methods for
the evaluation of the fermion determinant in
connection with path integrals~\cite{fuji}.

\section{Effective Chiral Lagrangians}

We consider the $SU(2)_L \bigotimes SU(2)_R$ lagrangian for $N_C$
coloured right- and left-handed quarks parame\-trized in the linear 
form
("$ \Sigma \,$- basis"), 
\begin{equation}\label{eq:L} 
{\cal L} = {\overline {\psi}}_L i {\not \! \partial} \psi_L
 + \overline{\psi}_R i {\not \! \partial} \psi_R
 -g (\overline{\psi}_L \Sigma \psi_R + 
      \overline{\psi}_R \Sigma^{\dagger} \psi_L )
 + {\cal L}_{boson}  ,
\end{equation}
with the $SU(2)$ matrix~\cite{donoh}
$\Sigma = \sigma + i {\vec{\tau}} {\cdot} \vec{\pi}$.
 The electromagnetic interaction is as usual, 
$\partial_\mu \rightarrow \partial_\mu + i e Q A_\mu$, where $A_\mu$
is the photon field and $Q = 1/2~ (1/3 + \tau_3)$. 
The explicit form of the boson  ~lagrangian
${\cal L}_{boson}$ is not needed.
Although in the strictly symmetric phase the constituent quark
mass $m$ has to vanish, we nevertheless introduce explicitly
a breaking term  
\begin{equation}\label{eq:Lm}
{\cal L}_m = - m \overline{\psi} \psi  .
\end{equation}
We treat $m$ as an explicit regularization parameter, which is finally
removed in
the symmetric phase by performing the limit $m \to 0$.
However, we do not break the chiral symmetry spontaneously by the 
standard redefinition of the scalar field $\sigma$. 

We change the quark fields~\cite{manm} by 
\begin{equation}\label{eq:transf} 
     {\psi_L}^{\prime} \equiv \xi^{\dagger} \psi_L ~~,
 ~~~~   {\psi_R}^{\prime} \equiv \xi \psi_R ,
\end{equation}
("$\xi$- basis")
defining a unitary matrix $\xi$ in such a way that the quark-pion coupling
(\ref{eq:L}) becomes replaced by a derivative coupling.
In the symmetric phase we obtain for nonvanishing $m$, 
after expanding in terms of the boson fields ($\sigma, \vec{\pi}$)
up to quadratic terms,
\begin{equation}\label{eq:xiexp} 
\xi \simeq
\exp [{ {i \vec{\tau} \cdot \vec{\pi}}
\over {2 (m/g + \sigma)}}]~, 
\end{equation}
which replaces the transformation matrix in the broken phase,
\begin{equation}\label{eq:xiexpT0} 
\xi = 
\exp [{ {i \vec{\tau} \cdot \vec{\pi}}
\over {2 F_{\pi}} }]~, 
\end{equation}
where $F_{\pi}$ is the pion decay constant.
In the chirally broken phase the transformation (\ref{eq:xiexpT0}) may be
obtained from (\ref{eq:xiexp}) by first shifting 
$\sigma \to <\sigma >+~ \sigma~ \hat{=} ~F_{\pi} + \sigma$
and then performing the well defined limit $m \to 0$, while 
the dynamical $\sigma$ field becomes heavy and  decouples.
However, approaching the symmetric phase, where $<\sigma >$ and $F_{\pi}$
vanish, the $\sigma$ field does not decouple and the mass parameter has to
be kept as $m\neq 0$.

The transformation (\ref{eq:transf}) yields 
\begin{equation}\label{eq:Lxi}
{\cal L} = {\overline {\psi}}^{\prime} (i {\not \!\! D} - m) \psi^{\prime}
 + {\cal L}_{boson}  
 + O (g \sigma \overline{\psi}^{\prime}  \psi^{\prime})~, 
\end{equation}
with 
\begin{equation}\label{eq:D}
 D_\mu = \partial_\mu +ie Q A_\mu
 + i {\overline V}_\mu + i {\overline A}_\mu \gamma_5 ~,
\end{equation} 
where e.g.
the axial current ${\overline A}_\mu$ becomes,
keeping  $m\neq 0$ fixed and expanding in the fields, 
\begin{equation}\label{eq:Aexpand} 
 {\overline A}_\mu \simeq \frac{1}{2} \left( \frac{g}{m} \right)^2 
[ ({m \over g} - \sigma) \vec{\tau} \cdot \partial_\mu \vec{\pi}
 - \vec{\tau} \cdot \vec{\pi} \partial_\mu \sigma ] + \cdots ~~.
\end{equation}
 It is important
to notice that because of the factor $({g / m})^2$ the current 
${\overline A}_\mu$  becomes singular for $m \to 0$, and as a consequence
the  symmetric limit is not immediately obtained.

Although the lagrangian remains invariant under the
transformation (\ref{eq:transf}), the jacobian is not unity,
 as it is well known. Thus, in order to obtain equivalent low-energy physics
in both representations, i.e. in the $"\Sigma \,$-" as well as in the
$"\xi \,$- basis", we  have to calculate  
the fermion determinant for the transformation (\ref{eq:transf}) 
at high temperature, and finally we have to perform the chirally symmetric
limit for $m \to 0$.

\section{Zeta- Regularization at High Temperature}

The jacobian for the transformation (\ref{eq:transf}) of the quark fields is 
determined by standard steps~\cite{donoh}.
First one defines infinitesimal transformations by introducing
a continous parameter $t$, $0 \le t \le 1$, and  extending the transformation
(\ref{eq:xiexp}), 
\begin{equation}\label{eq:xit} 
\xi \to \xi_{t} \equiv
\exp [{ {i t \vec{\tau} \cdot \vec{\pi}}
\over {2 (m/g + \sigma)}}]~ \equiv \exp [ i t \overline{\pi}]~.
\end{equation}
The effective lagrangian 
is defined by 
\begin{equation}\label{eq:WZW} 
 \int d^4x ~{\cal L}_{odd} \,
  \equiv 2 \int_0^1 dt ~ tr (\overline{\pi} \gamma_5) ~~,
\end{equation}
where  ${\cal L}_{odd}$ contains  the 
interactions in the odd-intrinsic-parity sector of the 
chiral model under consideration.
In the following we only consider local terms  as given by 
a Taylor expansion in $t$.

In order to  derive the anomalous coupling in leading order for
$\pi^0 \sigma ~\to 2 \gamma$ in the chirally symmetric phase,
we assume that the thermal bath is constituted
 by the $\pi^0$ and $\sigma$ fields,
to which external photons are coupled, i.e.  the photons are 
\underline{not} thermalized due to their small electromagnetic coupling.
In this respect we differ from 
Pisarski's treatment~\cite{pisa2}, where the photons are kept
 in thermal
equilibrium with $\pi^0$ and $\sigma$.

To be explicit we have to look for the
nonvanishing contributions to  
\begin{equation}\label{eq:L2ga1}
 {\cal L}_{T}^{2 \gamma} =
     {\lim_{m \to 0}  } ~ 
2 N_C {\int_0^1}~dt~tr^\prime~ \bigl( {{\tau_3 \pi^0} \over {2 (m/g + \sigma)}}
~\gamma_5 \, \zeta_s (x) \vert_{2\gamma} \bigr)~,
\end{equation} 
where  we apply the zeta function method~\cite{zeta} 
to regularize (\ref{eq:L2ga1}).
 $tr^{\prime}$ indicates the Dirac and flavour trace.
This method is combined with the expansion for the heat kernel~\cite{donoh} 
(for massive fermions) $H(x,\tau)$
by  the relation, 
\begin{equation}\label{eq:zeta} 
 \zeta_s (x) \equiv {1 \over {\Gamma(s)}}
  \int_0^{\infty} d\tau \tau^{s - 1} H (x, \tau) ~.
\end{equation}
At $T =0$, in the broken phase, 
 the $\zeta$-function regularization 
starts with a convergent expression  $\zeta_s (x)$
 for $s > 0$, which is then analytically continued to $s = 0$.
At high $T$, i.e. in the chirally symmetric phase,
we propose to continue in the variable $s$ in such a way
that a non-trivial limit for
 ${\cal L}_{odd}$, i.e. here for $ {\cal L}_{T}^{2 \gamma}$, 
 is finally obtained in the $m \to 0$ limit.

A few technical steps have to be performed:

\begin{itemize}

\item{} at high $T$ it is convenient to continue first
to euclidean coordinates and momenta ($p_E \equiv (\vec{p}, p_4)$);

\item{} introducing a complete set of energy (Matsubara) and 
momentum eigenstates one obtaines the expansion for the heat kernel
(\ref{eq:zeta}),  
\begin{equation}\label{eq:Hexpand} 
H (x, \tau)  \simeq  
 {{\tau^2} \over 2}
~ T~ \Sigma \int {{d^3p} \over {(2\pi)^3}} \exp [ - \tau (p_E^2 + m^2)]
~  \hat{\sigma}^2~.
\end{equation}
Here only the electromagnetic couplings in (\ref{eq:D})  are kept, where
expressed  in terms of the electromagnetic field tensor we have
$\hat{\sigma}   \equiv e Q\, \sigma_{\mu\nu} F^{\mu\nu} /2$.
 Terms of $O(\tau^0,\tau^1)$, which  either give vanishing contributions
after taking
the Dirac trace in (\ref{eq:L2ga1}),  
or do not contribute at $T = 0$, or lead to
interactions in higher orders of the boson fields at high $T$,
are dropped.
$\Sigma$ sums the fermionic frequencies $p_4 = 2 \pi T ( n + 1/2)$
with integer $n$;

\item{} in order to evaluate the Matsubara frequency sum we first
perform the $\tau$-integration in (\ref{eq:zeta})
inserting (\ref{eq:Hexpand}); it is convenient
 to introduce the following dimensionless functions~\cite{lands}
 $I(y^2; s + 2, {1 \over 2})$,
where $ y = m/ (2 \pi T)$.
For $s > 0$  they have the  series expansion\cite{lands},

\begin{eqnarray}\label{eq:Iexpand} 
 & &T ~  \Sigma  \int {{d^3p} \over {(2\pi)^3}} 
{ 1 \over {(p_E^2 + m^2)^{s + 2}}}
\equiv 
{{T^{ -2 s}} \over {(2 \pi)^{2s + 1}}}
I(y^2; s + 2, {1 \over 2}) \\ 
  = & &
{ {T^{-2s}} \over {(2\pi)^{2s+1}}}
{1 \over { 4 {\pi}^{3/2}\, \Gamma(s+2)}}
 ~\Sigma_0^{\infty} ~ {{(-)^k} \over {k!}} \Gamma(k+s+1/2)
  \zeta(2k + 2s + 1, {1 \over 2}) ~y^{2k}~ , \nonumber
\end{eqnarray}
where the Hurwitz zeta-function is related by
$\zeta (z, {1 \over 2}) \equiv (2^z - 1) \zeta (z)$ to the Riemann
function $\zeta (z)$.
It is crucial to realize that only
even powers in $(m/T)$ are present in (\ref{eq:Iexpand});

\item{} taking care of the proper dimension of $\tau$ in (\ref{eq:Hexpand})
 we multiply  by $T^{2s}$;  
an additional normalization factor $N(s)$,
with  $N(s = 0) = 1$, is allowed, which for $s \le 0$  is,
\begin{equation}\label{eq:N}
 N (s) =
      {{ (2 \pi)^{2s}~ \Gamma (1/2 -s) ~\zeta(1 - 2s, {1 \over 2})}
                   \over { \Gamma (3/2)~ \Gamma ( -s)}}~.
\end{equation}
\end{itemize}
 After performing these steps the leading term of the
 function $\zeta_s  \vert_{2\gamma}(x)$
 can be written as (in Minkowski metric) 
\begin{equation}\label{eq:zetaexpand} 
 \zeta_s \vert_{2\gamma} (x)  \simeq 
 {{ \pi N(s) \Gamma (s+ 2)} \over {(2 \pi)^{2s + 2}~ \Gamma (s)}}
   ~  I(y^2; s +2, {1 \over 2}) \, \hat{\sigma}^2 ~.
\end{equation}
We collect
the terms  in $\zeta_s \vert_{2\gamma} (x)$
 with different powers  in $m$,
i.e. in $y^2$, in order to finally perform the chirally symmetric limit.

\section{Anomalous Electromagnetic Couplings}

We now give the anomalous coupling  for
$\pi^0 \sigma ~\to 2 \gamma$ in the chirally symmetric phase.
Expanding (\ref{eq:L2ga1})
in the fields $(\sigma,\vec\pi)$, e.g.
$1/(m/g+\sigma) \simeq g/m- (g^2/m^2) \sigma$, it is seen, that
the symmetric limit $m\to 0$ is no longer trivial.
However, from (\ref{eq:zetaexpand})
 and the  expansion (\ref{eq:Iexpand}) with respect to $m^2$,
one finds 
 by adjusting the value of $s$ to $s=-1$  that the term proportional 
to $y^2$ indeed leads to a finite nonvanishing value in the $m \to 0$
limit.
In addition all the other coefficients in
(\ref{eq:Iexpand}) for $k\neq 1$ vanish
due to the pole at $s=-1$ of $\Gamma (s)$ in(\ref{eq:zetaexpand}),
from which we calculate 
\begin{equation}\label{eq:zetas-1} 
  \zeta_{s = -1} (x) =
 - \zeta(3,{1 \over 2})~y^2~\zeta_{s = 0} (x) =
 - {{ 7\,\zeta(3)} \over {32 \,\pi^2}} ~y^2~ \hat{\sigma}^2 ~.
\end{equation} 
Keeping the term linear in the $\sigma$ field the $m\to 0$ limit is
well defined, and 
the effective lagrangian for $\pi^0\sigma \to 2\gamma$ finally becomes 
\begin{equation}\label{eq:L2ga3}
 {\cal L}_{T}^{2 \gamma}
 =  -  {{ 7 \zeta (3)~ g^2 \alpha N_C} \over {96 \pi^3~T^2}}~ 
\epsilon^{\mu \nu \alpha \beta}~F_{\mu \nu}~F_{\alpha \beta}~
 ( \pi^0~\sigma )~.
\end{equation}
The effective photon coupling to neutral particles
$(\pi^0,\sigma)$ given in (\ref{eq:L2ga3}) is manifestly gauge invariant.
By comparing the above coupling at high $T$
 with the well known one~\cite{ra,rb} at $T = 0$  we note the substitution 
\begin{equation}\label{eq:FpiT}
 { 1 \over {F_\pi}} \rightarrow 
  { {7 \zeta (3) g^2} \over {4 \pi^2 T^2} }~\sigma~,
\end{equation}
which is in agreement with the derivation given by Pisarski~\cite{pisa1}.
%

In a corresponding fashion the anomalous couplings 
linear in $F_{\mu \nu}$ for e.g. $\gamma ~\to 3 \pi~ \sigma$
may be derived, which turn out to be in leading order
(keeping $s = - 2$)
proportional to $\zeta (5)~g^4/T^4$ at high temperature.

Generalizing the result  (\ref{eq:L2ga3})
to the case of $SU (3)$ allows to obtain the 
radiative two-photon couplings for 
the processes $\eta~(\eta^{\prime})~\sigma \rightarrow 2 \gamma$
in the chirally symmetric limit.
There are also additional anomalous couplings between photons
and mesons with more than one $\sigma$ field.

\section{Conclusions}

Finally, one may ask about phenomenological consequences,
i.e. possible  observable effects, of these high temperature
anomalous couplings.
An interesting case for testing (\ref{eq:L2ga3}) -
and those related to the anomalous interactions of vector
mesons~\cite{pisa1,pisa2}  at high $T$ - 
are photon- and dilepton measurements.

In this context we mention the production of
low-mass lepton pairs in heavy ion collisions at
 high energies~\cite{ceres}.
In order to interpret these data, especially for dilepton masses 
below  the $\rho$ meson mass,
a detailed knowledge of the electromagnetic interactions
of $\pi^0, \eta, \eta^{\prime}, \omega, \rho,$
etc. in a high density hadronic environment (i.e. maybe
in a chirally symmetric phase at high $T$)
is important.

Anomalous couplings of interacting bosons at high temperatures,
e.g. for $K K \to \pi \pi \pi \sigma$, have been
 derived~\cite{baier,pisa1,pisa2}. However, we do not find a simple
compact expression, comparable to the Wess-Zumino-Witten action~\cite{rg}
at $T = 0$, except that all these couplings
vanish in the $T \to \infty$ limit.

\section*{Acknowledgments}

We like to thank R.~D.~Pisarski for useful comments and  
discussions. R.~B. thanks the organizers H.~M.~Fried and B.~M\"uller,
and M.~Le~Bellac for inviting him to this Workshop.
Partial support by the EEC Programme "Human Capital
and Mobility", Network "Physics at High Energy Colliders",
Contract CHRX-CT93-0357
is acknowledged .
M.~D. is supported by Deutsche Forschungsgemeinschaft.


\begin{thebibliography}{99}
\bibitem{baier} R. Baier, M. Dirks, and O. Kober,
 \Journal{\PRD}{54}{2222}{1996}.
%
\bibitem{pisa1}
R. D. Pisarski, 
in {\em From Thermal Field Theory to Neural Networks:
a Day to Remember Tanguy Altherr}, eds. P. Aurenche et al.
(World Scientific Publishing, 1996), p. 41.
\bibitem{pisa2} 
R. D. Pisarski, \Journal{\PRL}{76}{3084}{1996}.
\bibitem{ra}
S. L. Adler, \Journal{\PR}{177}{2426}{1969};
J. S. Bell and R. Jackiw,  \Journal{\NCA}{60}{147}{1969}.
%
\bibitem{rb}
R. A. Bertlmann, {\em Anomalies in Quantum Field
 Theory} (Clarendon Press, Oxford, 1996),
and references therein.
%
\bibitem{rc}
L. Dolan and R. Jackiw, \Journal{\PRD}{9}{3320}{1974};
H. Itoyama and A. H. Mueller,
 \Journal{\NPB}{218}{349}{1983};
and references  in [1-3].
\bibitem{donoh}
J. F. Donoghue, E. Golowich, and B. R. Holstein,
{\em Dynamics of the Standard Model}
 (Cambridge Univ. Press, Cambridge, 1992).
\bibitem{manm}
A. Manohar and G. Moore,
 \Journal{\NPB}{243}{55}{1984}.
\bibitem{rj}
For a recent discussion of the QCD phase diagram
and applications of chiral symmetry at high temperature,
see R. D. Pisarski, lectures presented at
"Workshop on Finite Temperature QCD and Quark-Gluon Transport Theory",
Wuhan, PRC (April 1994).
\bibitem{fuji}
K. Fujikawa,
 \Journal{\PRD}{21}{2848}{1980};
 \Journal{\PRD}{29}{285}{1984}.
\bibitem{zeta}
N. K. Nielsen, M. T. Grisaru, H. R\"omer, and P.  van Nieuwenhuizen,
 \Journal{\NPB}{140}{477}{1978}; and references therein.
%
\bibitem{lands}
N. P. Landsman, 
 \Journal{\NPB}{322}{498}{1989}.
%
\bibitem{ceres}
CERES Collaboration, G. Agakichiev et al.,
 \Journal{\PRL}{75}{1272}{1995}.
\bibitem{rg}
J. Wess and B. Zumino, \Journal{\PLB}{37}{95}{1971};
E. Witten, \Journal{\NPB}{223}{422}{1983}.
\end{thebibliography}
\end{document}